\documentstyle[12pt,aps,epsf]{revtex}
\input{epsfig.sty}
\vspace{2cm}
\begin{document}
\title{\ \\ \ \\ \ \\ \ \\ \ \\  \ \\
CONDUCTANCE OF LUTTINGER-LIQUID WIRES CONNECTED TO RESERVOIRS}
\author{Dmitrii L. Maslov$^{(a,b,c)}$ and  Michael Stone$^{(a)}$}
\address{$^{(a)}$Department of Physics and $^{(b)}$Materials Research Laboratory\\
University of Illinois at Urbana-Champaign, Urbana, IL 61801-3080, USA}
\address{$^{(c)}$Institute for Microelectronics Technology, Academy of Sciences of Russia\\
Chernogolovka, 142423 Russia}
\maketitle
\thispagestyle{empty}
\bigskip

{\small We show that the dc conductance of a quantum wire containing a Luttinger
liquid and attached to noninteracting leads is given by $e^2/h$ per spin orientation,
regardless of the interactions in the wire. It is also shown that weak disorder
in the wire results in the temperature- or length-dependent corrections to the conductance.
The exponents of these dependences are determined by the interaction strength in the wire
and, in the leading order, are not affected by the presence of the non-interacting leads.
These results explain recent experiments on quasiballistic $GaAs$ quantum wires.}
\bigskip

\section{Introduction}
For non-interacting electrons, the conductance of narrow ballistic
 quantum wires connected to wide reservoirs is quantized in units of
$e^2/h$ \cite{REF:wees,REF:wharam}.  When the effects of interactions are
included this result is expected to be modified. In particular, when the
electrons in the wire form a one-dimensional
Luttinger liquid (LL) \cite{REF:haldane}, the
conductance is believed to be $Ke^2/h$ per spin orientation
\cite{REF:apel,REF:fisher-prb,REF:fukuyama}, where $K$ is the interaction dependent
parameter characterizing the LL.  For non-interacting
electrons $K=1$. For repulsive interactions $K<1$, and the conductance
should be reduced.

A recent experiment on very long $GaAs$ high mobility quantum wires
\cite{REF:tarucha} casts doubt on this picture.  It is known that the same
parameter $K$ enters the temperature dependence of the impurity
correction to the conductance, and using this the authors of
\cite{REF:tarucha} where able to estimate $K$ to be about $0.7$ for the
electron gas in their wires, implying a conductance reduction of $30\%$
in the ballistic limit. The actual reductions observed are only a few
percent of $e^2/h$ however.

\begin{figure}
\epsfig{figure=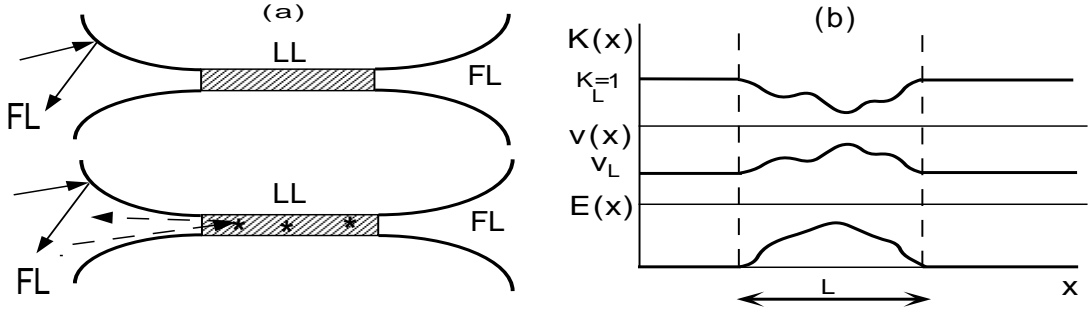,width=15 cm,height=15 cm, rheight=5 cm,rwidth=6 cm}
	      \caption{a)~Top: a narrow wire containing a Luttinger liquid and connected
to Fermi-liquid leads. Arrows show a typical electron trajectory giving rise to a  finite
conductance $g=e^2/h$. Bottom: the same for a wire with impurities
 (depicted by asterisks). Dashed arrows
show a typical trajectory giving rise to a correction to $e^2/h$. b)~Effective 1D model for the system
shown in (a) [top].}
	      \label{FIG:fig1}
	      \end{figure}

This observation has led us to re-assess the conventional analysis of the
charge transport in Luttinger wires. In this paper we will argue that
the conductance of a quantum wire attached to one-dimensional
non-interacting leads [which are intended to  model the
higher-dimensional Fermi-liquid (FL) reservoirs] is $e^2/h$ regardless of
the interactions in the wire itself\footnote{The result announced
 above was obtained by us \cite{REF:MS} simultaneously and independently with
Safi and Schulz \cite{REF:Safi_PRB}. By now, there are at least
 three more papers \cite{REF:Ponom,REF:Kawabata,REF:Fujimoto}
which,  each in its own way, confirm this result.}. This is because the finite
resistance of a ballistic wire is a {\it contact resistance}
\cite{REF:imry,REF:landauer,REF:buttiker,REF:glazman,REF:levinson} and comes entirely from processes
that take place outside the wire, where the electrons are not in the
LL state (cf. Fig.~\ref{FIG:fig1}a, top). 	

While it has been generally asserted in the literature that $g=Ke^2/h$ is the
correct result for interacting electrons
\cite{REF:apel,REF:fisher-prb,REF:fukuyama} (at least for wires longer than the
Fermi wavelength), we must remark that there have been previous comments
supporting the result $g=e^2/h$.  In their earlier paper
\cite{REF:fisher-prl}, Kane and Fisher remark that the ac conductance will
cross over to the non-interacting value at frequencies lower than
$\omega= v_F/L$, where $v_F$ is the Fermi velocity and  $L$ is the
wire length.  Matveev and Glazman \cite{REF:matveev} make a similar
remark when discussing the conductance of multi-mode interacting wires.

Having established the absence of conductance renormalization in pure wires, we will ask what would happen 
if we connect a weakly disordered wire to the leads.  In particular, we are interested in whether the temperature dependence of the conductance reflects the
presence of the LL-state in the wire or it is affected by the
presence of the leads as well. We will show that that the main contribution
to the temperature-dependent part of the conductance is determined by 
the local strength of interactions  in the wire and is the same as in the absence of the leads.
This result can be anticipated from the following
simple picture \cite{REF:Maslov}. Consider a weakly disordered wire containing an LL and adiabatically connected to the FL leads
(cf. Fig.~ \ref{FIG:fig1}a, bottom).
As has been said above, in the absence of disorder, the finite resistance
of a perfect wire ($=h/e^2$) is entirely due to contact resistance 
\cite{REF:imry,REF:landauer,REF:buttiker,REF:glazman,REF:levinson}: some of the electrons coming from 
the wide leads are reflected as the channel becomes narrower. This
reflection takes place outside the wire, where the electrons are in the 
FL state, therefore, the contact resistance is not affected by 
the interactions in the wire. Weak disorder in the wire 
gives rise to an additional contribution to the 
resistance. This contribution is determined by the scattering {\it in} the 
wire, where electrons are in the LL-state, and therefore this
contribution has features typical of an LL but not of an
FL.

These two results--the absence of conductance renormalization in perfect wires
and the temperature dependence of conductance determined by the local
strength of interactions in disordered wires--resolve the controversy 
around the experimental observations by Tarucha et al. \cite{REF:tarucha}
and suggest that these observations can be considered as an indication
of the LL-state in GaAs quantum wires.

This paper is organized as follows. In Sec.~II, we prove the
absence of conductance renormalization by using the equation of motion
for an LL in an external electric field. In Sec.~III, we 
obtain the same result by using the linear response theory. In Sec.~ IV, 
we find the temperature- and length-dependent corrections to the
conductance due to weak disorder in the wire. Our conclusions are given
in Sec.~V.

\section{LUTTINGER LIQUID UNDER EXTERNAL ELECTRIC FIELD: EQUATIONS OF MOTION}
We begin with replacing the original system containing an one-dimensional (1D) wire connected
 {\it adiabatically\/} to two-dimensional (2D) 
 leads (cf. Fig.~\ref{FIG:fig1}a) by the effective 1D system
 (cf. Fig.~\ref{FIG:fig1}b). 
This 1D system is an infinite LL which is described
 by the inhomogeneous effective 
interaction parameter $K(x)$. As is shown in  Fig.~~\ref{FIG:fig1}a, the interaction strength  starts from the
 value $K_{\rm L}=1$, e.g., in  the left lead,
goes through an arbitrary variation in the middle part
 of system (\lq\lq the wire\rq\rq) and approaches again the asymptotic value of $K_{\rm L}=1$ in the right lead. The 
asymptotic value $K_{\rm L}=1$ 
describes a non-interacting LL and reflects the fact that interactions in the leads
 result in  the formation of the FL-state, which for the
 present purposes can be viewed as the Fermi-gas state. Similarly, we allow for
 an arbitrary variation of the density wave velocity $v(x)$ in the wire 
 requiring only that it take asymptotic values $v_{\rm L}$ in the leads, which coincide with
 the Fermi-velocity of the non-interacting electrons. The electrostatic potential difference
 applied to the system produces largest electric field in the narrowest part of the system,
 i.e., in the wire. Accordingly, we assume that the electric field $E(x)$
 is zero outside the wire and undergoes
 an arbitrary variation in the wire.

The (real-time) bosonic action for a spinless LL is
\begin{equation}
S=\frac{\hbar}{2}\int d^2x \frac 1{K(x)}\left\{v(x) (\partial_x
\phi)^2-\frac 1 {v(x)} (\partial_t \phi)^2\right\}.
\label{(EQ:action_r)}
\end{equation}
We have chosen to  normalize the $\phi$ field  so that density of the 
electrons (minus the background density) and the (number) current
are given by
\begin{equation}
\rho=\partial_x \phi/\sqrt{\pi}, \qquad j=-\partial_t
\phi/\sqrt{\pi}.
\label{(EQ:rho_j)}
\end{equation}
The interaction with an external electromagnetic field $A_\mu$ is 
given by (the charge of the electron is taken to be $-e$)
\begin{equation}
S_{int}=\frac{e}{\sqrt{\pi}}\int d^2x \left\{A_0\partial_x \phi 
-A_1\partial_t\phi\right\},
\label{EQ:s_int}
\end{equation}
so the  equation of motion for the field (classical or quantum) is
\begin{equation}
\partial_t\left (\frac 1 {Kv}\partial_t \phi\right)
-\partial_x\left(\frac v K \partial_x \phi\right)
=\frac{e}{\sqrt{\pi}\hbar} E(x,t),
\label{EQ:eq_m}
\end{equation}
where $E=-\partial_xA_0+\partial_t A_1$ is the electric field. We assume
that the electric field is switched on at $t=0$, so that $E(x,t)=0$ for $t<0$
and $E(x,t)=E(x)$ for $t\geq 0$. As we see, the problem reduces now to that
of determining the profile of an infinite elastic string under the external force.
In this language, $\phi(x,t)$ gives the displacement of the string at point $x$ and at time $t$, while the
number current $j=-\partial_{t} \phi/\sqrt{\pi}$ is proportional to the
vertical velocity of the string.

\begin{figure}
\epsfig{figure=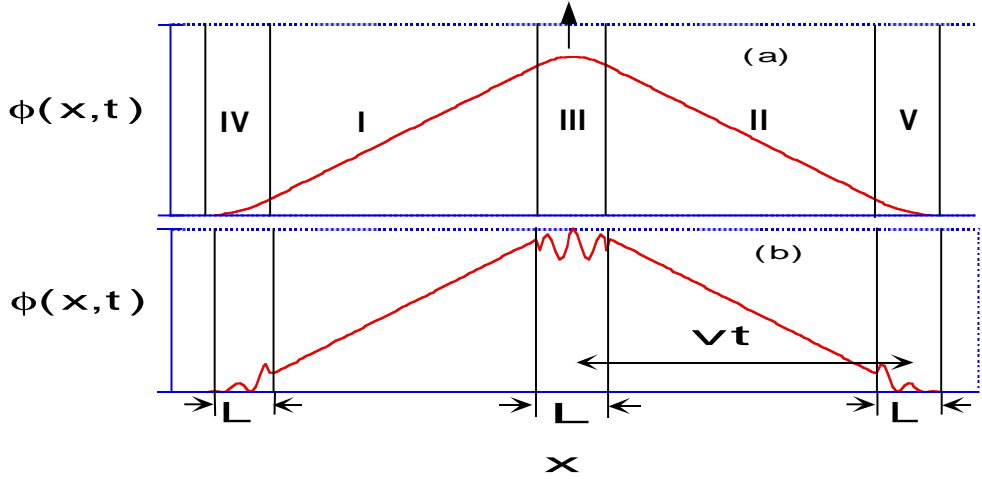,width=15 cm,rwidth=7 cm,rheight=6.5 cm}
\caption{ a)~Solution of the wave equation
in the homogeneous case
for $t=5L/v$. b)~Schematic solution in the inhomogeneous case for $t\gg L/v$.}
	      \label{FIG:string}
	      \end{figure}

To develop some intuition into the solution of Eq.~(\ref{EQ:eq_m}), we first solve
it in the homogeneous case, when $K=const$, $v=const$, and $E(x)=const$
for $|x|\leq 0$ and is equal to zero otherwise. In this case,  the solution
of Eq.~(\ref{EQ:eq_m}) for late times, i.e., when $t>L/v$, has the following form
\begin{equation}
\phi(x,t)=\frac{KeV}{2\sqrt{\pi}\hbar}
\cases{
t-\frac{x^2+L^2/4}{Lv}& for $|x|\leq L/2$;\label{EQ:hom}\cr
t-|x|/v& for $L/2\leq |x|\leq vt-L/2$\cr
\frac{v}{2L}\big(t-\frac{|x|-L/2}{v}\big)^2& for $vt-L/2\leq |x|\leq vt+L/2$\cr
0& for $|x|\geq vt+L/2$,
}
\end{equation}
where $V=EL$ is the total voltage drop. This solution is depicted in Fig.~\ref{FIG:string}a.
  The profile of the string consists
 of two segments (I and II in Fig.~\ref{FIG:string}a)
 whose widths--equal to $(vt-L)$--grow with time, and of three segments
  (III, IV and V in Fig.~\ref{FIG:string}a) whose widths are constant in time and equal to $L$.
In segments I and II, the  profile
 of the string $\phi(x,t)$ is linear in $x$, and therefore, being the solution
 of the wave equation, also in $t$; in segments III-V, the profile
  is parabolic.  Outside segments IV and V, the string is not perturbed yet and $\phi(x,t)=0$.
 As time goes on, the larger and
 larger part of the profile becomes linear. Thus, for late times,
 the motion of the string can be described as follows: the pulse
 produced by the force spreads outwards with the velocity $v$, involving
 the yet unperturbed parts of the string in motion; simultaneously,
 in all but narrow segments  in the middle and at the heads of the pulse, the string   
moves upwards with the $t$- and $x$-independent \lq\lq velocity\rq\rq\
 $\partial_{t}\phi=KeV/2\sqrt{\pi}\hbar$.
 In terms of the original transport problem, it means that the charge
 current $I=-ej$  is constant outside the wire (but not too close to the edges of
 the regions of where the electron density is not yet perturbed by the
 electric field) and given by $I=Ke^2V/h$. Therefore, the conductance
 (per spin orientation) is $g=Ke^2/h$. 

We now turn to the inhomogeneous case. As in the homogeneous case,
 the profile consists of several characteristic segments
 (cf. Fig.~\ref{FIG:string}b). In segments III-V, the profile is affected by the inhomogeneities in
 $K(x)$, $v(x)$, and $E(x)$ and depends on the particular choice of
 the $x$-dependences in all these quantities. In segments I and II however, the profile, being the solution of the
 free wave equation, is again linear in $x$ (and in $t$).
Requiring the slopes of the string be equal and opposite in segments I and II
(which is consistent with the condition of the current conservation),
 the solution in these regions can be written as $\phi(x,t)=A(t-|x|/v_{\rm L})$. 
The constant $A$ can be found by integrating
Eq.~(\ref{EQ:eq_m}) between two symmetric points $\pm a$ chosen outside the wire
\begin{equation}
-\int^{+a}_{-a}dx\partial_x\Big(\frac{v}{K}\partial_x\phi\Big)
=\frac{e}{\sqrt{\pi}\hbar}\int^{+a}_{-a}dxE(x)=\frac{eV}{\sqrt{\pi}\hbar}.
\label{EQ:cond_interm}
\end{equation}
Outside the wire, $K(x)=K_{\rm L}$ and $v(x)=v_{\rm L}$, thus $A=K_{\rm L}eV/2\sqrt{\pi}\hbar$.
Calculating the current, we get  $g=K_{\rm L}e^2/h$ and, recalling that $K_{\rm L}=1$, we finally arrive at $g=e^2/h$.
{\it Thus, the conductance is not renormalized by the interactions in the wire.}
The reason for this is that at late times ($t\gg L/v$)
the motion of the major part of the string
is not affected by the local inhomogeneities in the parameters $K$, $v$, and $L$.
It is this late-behavior of the solution, which one needs to determine the
{\it dc} conductance\footnote{Safi and Schulz \cite{REF:Safi_PRB} have shown that
the {\it ac} conductance is renormalized by the interactions in the wire in a non-trivial way.}.

\section{Kubo formula: perfect wire}
We are now going to derive the result that $g=e^2/h$ regardless of the interactions
in the wire in a different way, by using the Kubo formula for conductivity.
To do so, we adopt a simple model, in which the interaction parameter $K$
changes stepwise from the value $K_{\rm L}=1$ in the leads (for $|x|>L/2$) to the value $K_{\rm W}$ in the wire
(for $|x|\leq L/2$). The electric field is assumed to be zero outside the wire but
can take an arbitrary variation in the wire.

The charge
current $I$ is related to the field by
\begin{equation}
I(x,t)=\int^{L/2}_{-L/2}dx'\int \frac{d\omega}{2\pi}
 e^{-i\omega t}\sigma_{\omega}(x,x'){\bar E}_{\omega}(x'),
\label{EQ:current}
\end{equation}
where ${\bar E}_\omega(x)$ is the temporal Fourier component of the 
electric field
and $\sigma_{\omega}(x,x')$ is the non-local ac conductivity.
 In the 
Matsubara 
representation, $\sigma_{\omega}(x,x')$ is expressed via the 
(imaginary time) current-current correlation
function by the usual Kubo formula
\begin{equation}
\sigma_{\omega}(x,x')=\frac{ie^2}{\hbar
\omega}\int^{\beta}_{0}d\tau\langle
T_{\tau}^{*} j(x,\tau)j(x',0)\rangle
 e^{-i{\bar \omega}\tau}\Big|_{{\bar\omega}=i\omega
-\epsilon}.
\label{EQ:kubo}\end{equation}
In the bosonized
form, the particle-number current is given by $j=-i\partial_{\tau}\phi/\sqrt{\pi}$
and Eq.~(\ref{EQ:kubo}) reduces to
\begin{mathletters}
\begin{eqnarray}
\sigma_{\omega}(x,x')&=&\frac{e^2}{\hbar}\frac{i{\bar\omega}^2}{\pi\omega}G_{{\bar 
\omega}}^{0}(x,x')|_{{\bar\omega}\to i\omega-\epsilon},
\label{EQ:kubo1}\\
G_{{\bar \omega}}^{0}(x,x')&=&\int^{\beta}_{0}d\tau\langle 
T_{\tau}^{*} \phi(x,\tau)\phi(x',0)\rangle e^{-i{\bar \omega}\tau},
\label{EQ:prop_def}\end{eqnarray}
\end{mathletters}where $G^{0}_{{\bar \omega}}$ is the 
propagator of the boson field $\phi$. 
The propagator
$G^{0}_{{\bar \omega}}(x,x')$ satisfies the equation 
\begin{equation}
\Big\{-\partial_x\Big(\frac{v(x)}{K(x)}\partial_x\Big)
+{\bar \omega}^2\Big\}G^{0}_{{\bar \omega}}(x,x')=\delta(x-x'),
\end{equation}\label{EQ:prop}
complemented by the following boundary conditions: i) $G^{0}_{{\bar \omega}}(x,x')$ 
is continuous
at $x=\pm L/2$ and $x=x'$, ii) $\big[v(x)/K(x)\big]\partial_xG^{0}_{{\bar \omega}}(x,x')$ is
continuous at $x=\pm L/2$ but iii) undergoes
a jump of unit height at $x=x'$, {\it i.e.},
\begin{equation}
-\frac{v(x)}{K(x)}\partial_xG^{0}_{{\bar \omega}}(x,x')\Big|^{x=x'+0}_{x=x'-
0}=1.
\label{EQ:jump}\end{equation}
In addition, we assume that the infinitesimal dissipation is present in
the leads, so that $G^{0}_{{\bar \omega}}(\pm\infty,x')=0$. (In a real-
time formulation this corresponds to outgoing wave boundary
conditions.) We are interested in the
 dc limit of $\sigma_{\omega}(x,x')$ for which we need
to determine the behavior of $G^{0}_{\omega}(x,x')$ for $\omega\to 0$.
Note that according to Eq.~(\ref{EQ:current}),
one needs to know $G^{0}_{{\bar \omega}}(x,x')$ only for
 $|x'|\leq L/2$. In this region, the low-frequency
asymptotic form of the propagator can readily be shown
 to be independent of $x,x'$ and given by
$G^{0}_{\omega}(x,x')\to iK_{\rm L}/2\omega$ \cite{REF:MS}. Consequently, Eq.~(\ref{EQ:kubo1}) 
gives 
 \begin{equation}
 \lim_{{\omega}\to 0}\sigma_{\omega}(x,x')=\frac{K_{\rm L}e^2}{h}.
 \label{EQ:dccond}\end{equation}
 For a static electric field, ${\bar 
E}_{\omega}(x)=2\pi\delta(\omega)E(x)$, and Eq.~(\ref{EQ:current}) 
gives the $x$- and $t$-independent current
 \begin{equation}
 I=\frac {K_{\rm L}e^2}{h}\int^{L}_{0}dx'E(x')=\frac{K_{\rm L}e^2}{h}V,
 \label{EQ:dccurrent}\end{equation}
 {from} which we see that the conductance is 
 \begin{equation}
 g=K_{L}\frac{e^2}{h}.
\label{EQ:conduct}\end{equation}
Thus the conductance is 
determined by
the value of $K_{\rm L}$ in the leads  and does not depend on the value of
$K_{\rm W}$ in the wire. Recalling that $K_{\rm L}=1$, we get
$g=e^2/h$. We have thus confirmed the result obtained
in Sec.~II.

Notice that in this calculation we have implicitly taken the limit
$\omega \to 0$ while keeping $q$  finite
\cite{REF:lee-fisher,REF:fenton,REF:schulz}. The traditional order of limits is
$q\to 0$ before $\omega \to 0$. The latter yields the Drude formula,
which  has a divergent dc limit for perfect systems. The former
produces a finite (two terminal Landauer-B\"uttiker)  dc conductance even for a perfect system. 
It furthermore corresponds to the experimental situation where a static
field is applied over a finite region.
\section{Kubo formula: weakly disordered wire}
We now consider the case when a weak disorder potential $V(x)$ is present in the wire.
The action now acquires the term describing the backscaterring by disorder
\begin{equation}
\label{EQ:act_i}
S_{\rm i}=\frac{2}{a}\int \! dx \!\int^{\beta}_{0}\!d\tau V(x)
\cos (2k_Fx+2\sqrt{\pi}\phi),
\end{equation}
in which $a$ is the microscopic length cut-off. 
In what follows, we will 
determine the  leading
disorder-induced corrections to $\sigma_\omega$ directly
from the Kubo formula (\ref{EQ:kubo}) via
perturbation theory in $S_{\rm i}$ \footnote{The assumption of weak disorder 
is adequate for the experimental 
situation of Ref.~\cite{REF:tarucha}, where $L$ was at least $\simeq 6$
times smaller than the elastic mean free path of the unbounded 2DEG and the 
total observed change in the conductance was $1-5\%$ of $e^2/h$ depending 
on the wire length.}.
The first non-vanishing correction to the ensemble-averaged propagator
is given by
\begin{eqnarray}
\overline {\delta G(X,X')}=\frac{1}{a^2}\int dX_1dX_2\overline{V(x_1)V(x_2)}
\cos(2k_{\rm F}(x_1-x_2)\nonumber\\
\left\{\langle \phi(X)\phi(X')Q(X_1,X_2)\rangle_0\!-\!
\langle \phi(X)\phi(X')\rangle_0\langle Q(X_1,X_2)\rangle_0\right\}
\label{EQ:prop_corr}
\end{eqnarray}
where $X\equiv \{x,\tau\}$, $\overline{\mathstrut\dots}$
stands for ensemble averaging over disorder realizations,
 $\langle\dots\rangle_0$ for thermodynamic averaging and
\begin{equation}
Q (X_1,X_2)=e^{i2\sqrt{\pi}\left[\phi(X_1)-
\phi(X_2)\right]}
\label{EQ:gamma}
\end{equation}
At this stage, for the sake of simplicity, we choose $V(x)$ 
in the form of white-noise: $\overline{V(x)}=0$ and $\overline{V(x_1)V(x_2)}=
n_{\rm i}u^2\delta(x_1-x_2)$, where $n_{\rm i}$ is the concentration
of \lq\lq impurities\rq\rq, and $u$ is the \lq\lq impurity\rq\rq\
strength. The effective elastic mean free path $\ell$ (in the absence of the
interactions) can then be defined as
 $1/\ell=n_{\rm i}u^2/a^2\omega_{\rm F}^2$,
where $\omega_{\rm F}$ is the (non-universal) energy ultraviolet cut-off
(of the order of the Fermi energy)\footnote{Although in the real $GaAs$ system
the impurity potential is long-ranged, the expectation is that this
simplification is not going to affect the final results significantly,
as soon as $\ell$ is replaced by the correct mean free path 
for a more realistic potential.}.
We then obtain
the correction to the non-local conductivity
\begin{eqnarray}
-\overline{\delta\sigma_{\omega}(x,x')}=\frac{2ie^2 {\bar\omega}^2
\omega_{\rm F}^2}{\pi\ell\omega}
\int^{L/2}_{-L/2}d{\bar x}G^{0}_{ {\bar\omega}}(x,\bar x)G^{0}_{ {\bar\omega}}(\bar x,x')\nonumber\\
\left[F_{0}(\bar x)-F_{ {\bar\omega}}(\bar x)\right]|_{ {\bar\omega}\to i\omega-\epsilon},
\label{EQ:cond_corr}
\end{eqnarray}
where $F_{ {\bar\omega}}(x)$ is the $\tau$-Fourier transform of the (inhomogeneous) $2k_F$ 
density-density correlation function 
\begin{eqnarray}
&F(x,\tau)=\langle Q(x\tau,x0)\rangle_0=\nonumber\\
&\exp\Big[-\frac{4\pi}{\beta}
\sum_{ {\bar\omega}}(1-e^{-i {\bar\omega}\tau})G^{0}_{ {\bar\omega}}(x,x)\!\Big].
\label{EQ:F}
\end{eqnarray}
To evaluate the function $F$ in Eq.~(\ref{EQ:F}), one needs to know 
$G^{0}_{ {\bar\omega}}(x,x')$ only for $-L/2\leq x=x'\leq L/2$. Straightforward, albeit 
lengthy, 
algebra leads to
\begin{eqnarray}
\!\!\!\!G^{0}_{ {\bar\omega}}(x,x)=\frac{K_{\rm W}}{2| {\bar\omega}|}\!+\!\frac{K_{\rm W}}{| {\bar\omega}|}\frac{\kappa_{-}^2e^{-L/L_{ {\bar\omega}}}+
\kappa_{+}
\kappa_{-}\cosh(2x/L_{ {\bar\omega}})}{e^{L/L_{ {\bar\omega}}}
\kappa_{+}^2-e^{-L/L_{ {\bar\omega}}}\kappa_{-}^2},
\label{EQ:prop_xx}
\end{eqnarray}
where $L_{ {\bar\omega}}=v_{\rm W}/| {\bar\omega}|$, $v_{\rm W}$ is the density-wave velocity in the
wire and $\kappa_{\pm}=1/K_{\rm W}\pm1/K_{\rm L}$. We now 
consider separately the cases of \lq\lq high\rq\rq\ 
($ v_{\rm W}/L\ll T\ll \omega_{\rm F}$)
and \lq\lq low\rq\rq\ ($T\ll v_{\rm W}/L$)
temperatures. (The quotations marks here are implied to mean that, say, 
\lq\lq low\rq\rq\ -$T$ case can be alternatively viewed as the \lq\lq 
long\rq\rq\ -$L$ 
case and vice versa.)

At \lq\lq high\rq\rq\ temperatures, the second
term in Eq.~(\ref{EQ:prop_xx}) is exponentially small [$\propto 
\exp(-(L-2x)/L_{ {\bar\omega}}$]  unless $ {\bar\omega}=0$. The term $ {\bar\omega}=0$ gives zero 
contribution to the sum in Eq.~(\ref{EQ:F}), as can be seen by performing
the infrared regularization of the propagator 
($| {\bar\omega}|\to\sqrt{| {\bar\omega}|^2+m^2}$) and then letting $m\to 0$. Thus, in this case
only the first term in Eq.~(\ref{EQ:prop_xx}) has to be taken into account. This term is precisely the same, however, as in the case of a homogeneous
Luttinger liquid with the parameter $K_{\rm W}$. Already at this stage the result 
that the $T$-scaling of the conductance is determined by $K_{\rm W}$ can be 
anticipated. The function $F$ is now $x$-independent and is given by
\begin{equation}
F=\Big[\frac{2\pi/\omega_F\beta}{\sin \pi\tau/\beta}\Big]^{2K_{\rm W}}
\label{EQ:F_high}
\end{equation}  
After analytic continuation  $ {\bar\omega}\to i\omega-\epsilon$, the limit
$\omega\to 0$ can be taken. As we saw in Sec.~III,  the remaining two propagators
in Eq.~(\ref{EQ:cond_corr}) are $x$- and $x'$ independent in this limit: $\lim_{\omega\to 
0}G^{0}_{\omega}(x,x')=iK_{\rm L}/2\omega$.  The rest of the calculations
proceeds exactly as in the homogeneous case.  Recalling the
result for the conductance of a perfect wire Eq.~(\ref{EQ:conduct}), the final result for
the conductance can be written as
\begin{equation}
\!\!\!\overline{g}=K_{\rm L}\frac{e^2}{h}\Big\{1\!\!-\!\!CK_{\rm L}\frac{L}{\ell}\Big[\frac
{\hbar\omega_{\rm F}}{2\pi T}\Big]^{2(1-K_{\rm W})}\Big\},
\label{EQ:res_high}
\end{equation}
where
\begin{equation}
C=8\sqrt{\pi}\sin(\pi K_{\rm W})\frac{\Gamma(1-K_{\rm W})}{\Gamma(\frac{1}{2}+K_{\rm W})}\Big[\Gamma(K_{\rm W})\Big]^2.
\label{EQ:const}
\end{equation}
Note that $K_{\rm L}$ enters only the prefactors, while the exponent of 
$T$-scaling
is determined by $K_{\rm W}$\footnote{Safi and Schulz \cite{REF:Safi_NATO} have found corrections to the leading $T$-dependent term in Eq.~(\ref{EQ:res_high}) which arise from additional scattering at the boundaries between disordered and perfect regions. These corrections become dominant only if the interaction is strongly attractive, which is not the case for the semiconductor heterostructures.} Tracing back the calculations, we can now see the reason
for this. The $K_{\rm L}$-dependence comes from the propagators
in the prefactor of Eq.~(\ref{EQ:cond_corr}) which depend on the  
frequency of the external field $\omega$. In the limit $\omega\to 0$, these propagators 
become long-ranged
and \lq\lq know\rq\rq\ only about $K_{\rm L}$ but not $K_{\rm W}$. The 
$T$-scaling comes the propagator entering the
function $F$. This propagator does not depend on $\omega$. In the 
\lq\lq\ high\rq\rq\-temperature limit, it becomes short-ranged and 
\lq\lq knows\rq\rq\ only about $K_{\rm W}$ but not $K_{\rm L}$.

In the case of \lq\lq low\rq\rq\ temperatures, the analysis is much more cumbersome.
Referring the reader for details to Ref.~\cite{REF:Maslov}, we give
only the final expression the real part of the conductance
\begin{equation}
\overline{Re 
g}=\frac{e^2}{h}-\frac{e^2}{h}2\pi^2\Gamma(K_{\rm W})\frac{L}{\ell}\Big[\frac{2L\hbar\omega_{\rm F}}
{ v_{\rm W}}\Big]^{2(1-K_{\rm W})}.
\label{EQ:res_low}
\end{equation}
Comparing this result with the analogous result for the
 homogeneous case \cite{REF:apel,REF:fukuyama}, we see that
the exponent of the length-dependence is the same as if the leads would
be absent\footnote{ For electrons with spin,
the exponent $2-2K_{\rm W}$ is replaced by $2-K^{\rho}_{\rm W}-K^{\sigma}_{\rm 
W}$ or, if the $SU(2)$ symmetry of the underlying Hubbard model is 
preserved, {\sl i.e.}, $K^{\sigma}_{\rm 
W}=1$, by $1-K^{\rho}_{\rm W}$, where $K^{\rho,\sigma}_{\rm W}$ are the 
parameters of the charge (spin) parts of the Luttinger-liquid in the wire.}.
The $L$-scaling is still symmetric to $T$-scaling upon replacement $T\Leftrightarrow v/L$ 
(apart from the additional factor of $L$ entering in the combination with the mean free
path). Contrary to the homogeneous case however, this symmetry exists only if the
leads are not interacting, i.e., $K_L=1$.
\section{Discussion and conclusions}
In Secs.~II and III, we have shown that the conductance of a perfect wire containing a Luttinger liquid and connected to noninteracting leads is not renormalized by the interactions in the wire. It remains at the noninteracting value $g=e^2/h$ per spin orientation. In Sec.~IV, we have shown that weak disorder in the wire leads to the
temperature- or length-dependent corrections to the conductance, the exponent of which being determined (in the leading order) by the interaction strength in the wire and independent of the presence of the leads.

These results are consistent with the recent experimental observations by Tarucha et al. \cite{REF:tarucha}, who observed the anomalous temperature dependence of the conductance at lower temperatures with exponent corresponding to $K_{\rho}\approx 0.7$
but no renormalization of conductance at higher temperatures. According to the previous theory, this value of $K_{\rho}$ extracted from the temperature dependence should have implied a 30\% renormalization of $g$ from the value of $e^2/h$ at higher temperatures, which clearly contradicts to the data. 

It is certainly tempting to ask whether the observed temperature dependence \cite{REF:tarucha} really comes from the Luttinger--liquid-like behavior
of the electrons in the wire or there exists some other plausible
explanation for this. One of the possible reasons for the $T$-dependence might be the phase-breaking scattering of weakly-localized electrons at phonons. The counter-argument to this assertion is that it would be difficult to explain
the anomalous $T$-dependence of the conductance observed in Ref.~\cite{REF:tarucha} ($-\delta g\propto T^{-0.3}$)
by the scattering of non-interacting electrons at three-dimensional phonons.

Even within the Luttinger-liquid model, the impurity scattering is not the only possible mechanism of the $T$-dependence. The other one is the scattering of electrons from the non-adiabatic openings of the wire into the leads\footnote{We thank A.~Gogolin for attracting our attention to this issue.}.
 One can imitate this mechanism in the 1D model by putting the point scatterers at the ends of the wire, the $S$-matrixes of these scatterers being equal to that of the non-adiabatic leads. The conductance of such a system is known to be $T$-dependent \cite{REF:fisher-prb,REF:gogolin}. However, the $T$-dependent correction to the conductance is length-independent in this mechanism, whereas the observed deviation of the conductance from $e^2/h$  \cite{REF:tarucha} is definitely more pronounced in longer wires. This gives more support
to the disorder mechanism considered in Sec~IV, although the coexistence of two mechanism is quite plausible. We hope to see more experiments in the nearest future providing more checks for the current and new theories.

Finally, we would like to mention that the result of the absence of conductance renormalization by the interactions
in a Luttinger-liquid wire discussed in this paper should {\it not} be applied to chiral Luttinger
liquids \cite{REF:Wen,REF:Moon} formed at the edges of fractional quantum Hall effect (FQHE) systems.
The two-terminal conductance of a FQHE system is $g_{FQHE}=\nu e^2/h$, where $\nu$ is the filling
factor. Despite the formal similarity of this expression with the conductance of
a homogeneous non-chiral Luttinger liquid ($=Ke^2/h$), the physical meanings of these
two quantities are quite different. In FQHE system, $g_{FQHE}$ is the Hall
(non-dissipative) conductance \cite{REF:Allan,REF:Moon}, whereas the conductance
of non-chiral Luttinger liquids is the  longitudinal (dissipative) one.
Moreover, the finiteness
of $g_{FQHE}$ is not due to the presence of reservoirs (as in the non-chiral case).
The \lq\lq renormalization\rq\rq\ of $g_{FQHE}$ from the value of $e^2/h$  is due
to the interactions in the bulk, and the bosonic description of a chiral
Luttinger liquid \cite{REF:Wen} is constructed in such a way that the correct result
for $g_{FQHE}$ is obtained.

\section{Acknowledgements}
This work was supported by the NSF of USA under Grants No. DMR89-20538 and DMR94-24511. We both thank D.~Loss and P.~M.~Goldbart for discussions and encouragement. DLM is grateful to Y.~B.~Levinson for numerous conversations on the physics behind the Landauer formula and to A.~Gogolin, M.~P.~A.~Fisher, G.~B.~Lesovik, A.~H.~Macdonald, K.~A.~Matveev, I.~Safi, N.~Sandler, H.~J.~Schulz, S. Tarucha, 
and A.~Yacoby for stimulating discussions on various issues touched upon in this paper. He also thanks organizers of the XXXI Rencontres de Moriond conference \lq\lq Correlated Fermions and Transport in Mesoscopic Systems\rq\rq\ for helping to make his participation in this conference possible and for providing a beatiful atmosphere in which many interesting discussions have taken place.

\begin{center}
CONDUCTANCE DANS UN LIQUIDE DE LUTTINGER\\ CONNECTE A DES RESERVOIRS\\
\bigskip
Dmitrii L. Maslov et Michael Stone
\end{center}
{\small Nous montrons que la conductance continue 
dans un fil quantique qui contient un liquide de Luttinger 
et qui est connect\' e \` a des r\' eservoirs non-interactifs
est donn\' ee par $e^2/h$ pour chaque orientation de spin.
Ce r\' esultat est ind\' ependant de l'interaction dans le fil.
Il est aussi montr\' e que la pr\' esence d'un faible d\' esordre
dans le fil r\' esulte en des corrections de la conductance
d\' ependant de la temp\' erature et de la longueur du fil.
Les exposants de ces corrections sont d\' etermin\' es
par la force de l'interaction dans le fil, et, au premier ordre,
ne d\' ependent pas de la pr\' esence des r\' eservoirs non-interactifs.
Ces r\' esultats expliquent des exp\' eriences r\' ecentes sur des
fils quantique de $GaAs$ dans le r\' egime quasi-ballistique.}

\end{document}